\documentclass[prl,twocolumn,amsmath,amssymb,floatfix]{revtex4}
\usepackage{graphicx}
\usepackage{bm}
\usepackage{bbold} 

\newcommand{\bra}{\langle}
\newcommand{\ket}{\rangle}
\newcommand{\sgn}{\mathrm{sgn}}

\newcommand{\eref}[1]{Eq.~\ref{#1}}

\newcommand{\fref}[1]{Fig.~\ref{#1}}

\newcommand{\rcite}[1]{Ref.~\onlinecite{#1}}

\newcommand{\nn}{\nonumber}
\newcommand{\be}{\begin{equation}}
\newcommand{\ee}{\end{equation}}
\newcommand{\beq}{\begin{eqnarray}}
\newcommand{\eeq}{\end{eqnarray}}

\begin{document}

\title{Measuring Berry curvature with quantum Monte Carlo}
\author{Michael Kolodrubetz}
\affiliation{Physics Department, Boston University, 590 Commonwealth
Ave., Boston, MA 02215}

\begin{abstract}
The Berry curvature and its descendant, the Berry phase, play an important
role in quantum mechanics.  They can be used to understand the Aharonov-Bohm
effect, define topological Chern numbers, and generally to investigate
the geometric properties of a quantum ground state manifold.   While Berry curvature
has been well-studied in the regimes of few-body physics and non-interacting
particles, its use in the regime of strong interactions is hindered by the lack of
numerical methods to solve it.
In this paper we fill this gap by implementing a quantum Monte Carlo method to solve for the Berry 
curvature, based on interpreting Berry curvature as a leading correction to 
imaginary time ramps.  We demonstrate our algorithm using the transverse-field Ising model in one
and two dimensions, the latter of which is non-integrable. Despite 
the fact that the Berry curvature gives information about the phase of the 
wave function, we show that our algorithm has no sign or phase problem for
standard sign-problem-free Hamiltonians.  Our algorithm
scales similarly to conventional methods as a function of system size and energy gap, and therefore should
prove a valuable tool in investigating the quantum geometry of many-body systems.
\end{abstract}

\maketitle

From the Aharonov-Bohm effect \cite{Berry1984_1,Aharonov1987_1} 
to topological insulators 
\cite{Haldane1988_1,Kane2005_1,Bernevig2006_1,Konig2007_1,Fu2007_1,Hsieh2008_1}, 
geometry and topology play a major role in modern condensed matter physics. 
Topological properties of such systems yield
edge states, quantized transport, and other robust physical quantities
\cite{Berry1984_1,Thouless1982_1,Niu1985_1,Hatsugai1993_1,Canali2003_1,Kane2005_1,Sheng2006_1,Bernevig2006_1}.
Topologically non-trivial systems have even been proposed as having 
major implications for cosmology \cite{Levin2005_1}.

Nearly all of the topological invariants in quantum mechanical systems are
based on the concept of geometrical phase, a.k.a. Berry phase, a quantity
that reflects the geometry of the ground state manifold \cite{Berry1984_1}.  Berry phase
is directly tied to a local tensor known as the Berry curvature.
Integrals of the Berry curvature define many important
topological quantities such as the Chern number \cite{Thouless1982_1,Hatsugai1993_1}
and its presence in electron systems plays a role in the 
anomalous quantum Hall effect \cite{Haldane2004_1} 
and crystal polarization \cite{KingSmith1993_1,Resta2000_1}.

While the Berry curvature and its integrals are relatively well understood 
for non-interacting electrons \cite{Thouless1982_1,Niu1985_1,Kane2005_1,Sheng2006_1}, 
their use in strongly
interacting systems remains in its infancy.  For weakly-correlated systems,
density functional theory does very well \cite{KingSmith1993_1}, but for
strongly-correlated systems the best exact method to determine the
Chern number is currently numerical diagonalization \cite{Hafezi2008_1,Resta2000_1},
which scales very poorly with system size.  Therefore, it is important to develop methods
to extend calculations of the Berry curvature to larger system sizes.  

In this paper, we develop a quantum Monte Carlo (QMC) method for measuring
the (many-body) Berry curvature.  QMC methods remain the gold
standard for exact numerical methods in many-body physics, 
scaling efficiently with system size for a wide class of problems 
\cite{Hammond1994_1,Nightingale1999_1,Anderson2007_1}.
Here we described and implement such a method which, 
similar to the original work by Berry \cite{Berry1984_1}, 
uses spin systems; in particular, we demonstrate our ideas using the 
quantum spin-1/2 transverse-field Ising (TFI) model in $d$ dimensions 
\cite{Lieb1961_1,Pfeuty1971_1,Jongh1998_1,Hamer2000_1,Liu2013_1},
which is a non-integrable strongly-interacting spin system for $d\geq 2$.
The algorithm can be readily extended to other sign-problem-free spin
\cite{Syljuasen2002_1,Sandvik2007_1,Melko2008_1}, 
bosonic \cite{Ceperley1995_1,Clark2006_1,Boninsegni2012_1},
or even certain fermionic systems \cite{Sugiyama1986_1,Sorella1989_1}.

Given that the Berry curvature is a measure of the invariant ground state phase, 
it is surprising that QMC methods can solve
this quantity without encountering the notorious sign problem.
To accomplish this, we rewrite the Berry curvature as the leading-order correction to an
asymmetric ramp in imaginary time \cite{Gritsev2012_1,Liu2013_1}, whose dynamics 
can be solved exactly using sign-free QMC methods.  At the end of the paper,
we show that this algorithm scales with system size $L$ and gap $\Delta$
comparably to ground state algorithms for more 
conventional observables, demonstrating that use of our algorithm on
large and complicated systems is quite feasible.

\emph{Berry curvature from ramps -- }
Consider an arbitrary manifold of Hamiltonians, $H(\vec \lambda)$, 
parameterized by some externally-controlled parameters $\vec \lambda$ (e.g.,
magnetic field).  Given the ground states $|\psi_0(\vec \lambda)\ket$,
the Berry phase is defined for some 
closed loop $\mathcal C$ in parameter space as follows. Consider ramping the 
parameters around this loop adiabatically and returning to the initial point
$\vec \lambda_i$. A naive expectation is that the wave function
will return to $|\psi_0(\vec \lambda_i)\ket$ up to a dynamical
phase factor $e^{i\gamma_\mathrm{dyn}}$, where $\gamma_\mathrm{dyn} = -\int E_0 dt$.
However, there is an additional phase factor $\gamma_B$ known as the geometric
or Berry phase \cite{Berry1984_1}, which derives from the quantum geometry of the ground state 
manifold.  The Berry phase is given by
\be
\gamma_B = i \oint_{\mathcal C} \bra \psi_0 | \vec \nabla | \psi_0 \ket \cdot d \vec \lambda
\equiv  \oint_{\mathcal C} \vec A \cdot d \vec \lambda ~,
\ee
where $\vec A$ is the Berry connection.  If we think of $\vec A$ as a magnetic
vector potential, then its curl is the ``magnetic field'' $F$, called the Berry curvature:
\be
F_{\mu \nu}= \partial_\mu A_\nu - \partial_\nu A_\mu ~.
\ee

Just as the phase of a charged particle acquires an Aharonov-Bohm phase
when moving around a magnetic flux, the surface
integral of the Berry curvature over a manifold $\mathcal M$ with $\mathcal C$
as its boundary gives the Berry phase: $\gamma_B = \int_\mathcal{M} F\cdot dS$. If
$\mathcal M$ is a closed manifold, then single-valuedness of the wave
function demands that the phase be $2\pi n$ for some integer  
\be
n = \frac{1}{2\pi} \oint_\mathcal{M} F \cdot dS ~,
\ee
where $n$ is a topological invariant known as the first Chern number.
In the language of effective magnetic field, $n$ is the number of flux
quanta piercing the surface $\mathcal M$, which is an integer due to Dirac 
monopole quantization \cite{Dirac1931_1}.
It can be used to define the topological order of many materials,
e.g., quantum Hall systems and topological insulators 
\cite{Thouless1982_1,Hatsugai1993_1,Kane2005_1,Sheng2006_1}.
Therefore, the Berry curvature plays a crucial role in the geometry and topology
of the ground state manifold.

Recently, Gritsev and Polkovnikov \cite{Gritsev2012_1} pointed out that the Berry curvature emerges
naturally as a leading order corrections to adiabatic dynamics in ramped systems.  While
Monte Carlo methods are unable to simulate such real-time dynamics, a similar analysis in
imaginary time \cite{Liu2013_1} gives the Berry curvature for asymmetric ramps.
More explicitly, consider an imaginary time ramp along some direction $\lambda_\mu$ at rate
$v_\mu =d\lambda_\mu / dt$.  Then, stopping at a fixed point $\vec \lambda_f$ in parameter space, the wave function
to lowest order in $v_\mu$ is given by \cite{DeGrandi2011_1}
\be
|\psi(v_\mu)\ket \approx |0\ket - v_\mu \sum_{n\neq 0} |n\ket 
\frac{\bra n | \partial_\mu H | 0 \ket}{(E_n-E_0)^2} + O(v_\mu^2) ~,
\ee
where $|n\ket$ labels the energy eigenstates of $H(\vec \lambda_f)$ with 
energy $E_n$ and non-degenerate ground state $|0\ket$.
Now imagine propagating the bra and ket asymmetrically, and taking the 
matrix element of the generalized force $\partial_\nu H$.  Noting that $\bra \psi(-v_\mu) | \psi(v_\mu) \ket=1$
to order $v_\mu^2$, we find the leading contribution to this overlap is
\be
\frac{\bra \psi(-v_\mu) | \partial_\nu H | \psi(v_\mu) \ket}{\bra \psi(-v_\mu) | \psi(v_\mu) \ket}
\approx \bra 0 | \partial_\nu H | 0 \ket - i v_\mu F_{\mu \nu} + O(v_\mu^2)~,
\label{eq:linearF}
\ee
where $F_{\mu\nu}$ is the (many-body) Berry curvature
\footnote{To show this, we use the form of the Berry curvature in terms 
of generalized forces: $F_{\mu \nu} = i \sum_{n \neq 0} 
\frac{\bra 0 | \partial_\mu H | n \ket \bra n | \partial_\nu H | 0 \ket - 
(\mu \leftrightarrow \nu)}{(E_n - E_0)^2}$ \cite{Berry1984_1}.}. 
Since $\partial_\nu H$ is Hermitian, the first term
in \eref{eq:linearF} is real, while the 
second term is strictly imaginary.  Thus,
\beq
\nn
 v_\mu F_{\mu \nu} &\approx& -\mathrm{Im}\left[ \frac{\bra \psi(-v_\mu) | \partial_\nu H | \psi(v_\mu) \ket}
{\bra \psi(-v_\mu) | \psi(v_\mu) \ket} \right] 
\\&=& 
\mathrm{Re}\left[ \frac{\bra \psi(-v_\mu) | \big( i \partial_\nu H \big) | \psi(v_\mu) \ket}
{\bra \psi(-v_\mu) | \psi(v_\mu) \ket} \right] ~.
\label{eq:Fmunu_basic}
\eeq

By taking the slow limit of these asymmetric ramps and
measuring the generalized force $i\partial_\nu H$, we can extract the Berry curvature.  
These imaginary time ramps are amenable to QMC methods, as long
as the Hamiltonian is sign free for all values of $\lambda_\mu$ during the ramp.  
As a demonstration of this method, we now 
construct an algorithm for computing \eref{eq:Fmunu_basic} in the TFI model
using an extension of the quasi-adiabatic QMC method \cite{Liu2013_1}.

\emph{Application to the TFI model -- }
Consider the TFI model on a $d$-dimensional lattice, with Hamiltonian
\be
H=-J \sum_{<jj'>} \sigma^x_j \sigma^x_{j'} - h \sum_{j} \sigma^z_j ~, 
\label{eq:Hising_basic}
\ee
where $J>0$ is the ferromagnetic Ising interaction acting on nearest
neighbors $j$ and $j'$, $h$ is the transverse
field, and $\sigma^{x,y,z}$ are Pauli matrices.  The ground state of this 
Hamiltonian has a quantum phase transition from paramagnet to ferromagnet at
the critical point $h_c$, where $h_c = \pm J$ in one dimension \cite{Lieb1961_1} and 
$h_c \approx \pm 3.04458 J$ in two dimensions \cite{Liu2013_1}.
To get non-zero Berry curvature, we introduce
a third parameter $\phi$ corresponding to a global rotation of
all the spins by an angle $\phi/2$ about the $z$-axis.  The Berry
phase of this extended TFI model has been investigated experimentally \cite{Peng2010_1}
and theoretically \cite{Zhu2006_1,Kolodrubetz2013_2} in the integrable one-dimensional case.  Here, we
numerically extend this analysis to arbitrary dimensionality $d$ by 
using QMC methods.

To fix the overall energy scale, we reparameterize the couplings as
$h=s$ and $J=1-s$.  Then the Hamiltonian described above can be written
\be
H(s,\phi) = -s \sum_j \sigma^x_j - (1-s)\sum_{<j j'>} 
\big[ \sigma^+_j \sigma^-_{j'} + e^{i \phi} \sigma^+_j \sigma^+_{j'} + \mathrm{h.c.}] ~,
\label{eq:Hsphi}
\ee
where $\sigma^{\pm} = \frac{1}{2} (\sigma^z \pm i \sigma^y)$.  We are interested in
the non-trivial component $F_{s\phi}$ of the Berry curvature tensor, which 
is a function of the tuning parameter $s$.
For this choice of parameters, consider a ramp from $s=0$ to $s=1$.
At $s=1$, the ground state consists of all spins pointing to the right,
which we denote $|\Rightarrow \ket$. Similarly, at $s=0$, the ground state consists
of either all states pointing up or all down; we manually break the
symmetry by choosing $|\Uparrow\ket$
\footnote{Note that, by breaking the $\mathbb Z_2$ symmetry, we
expect to measure the Berry curvature with respect to the ground state manifold
with the chosen magnetization (in the limit of large system size, i.e.,
exponentially small gap between ground state sectors).  This is important because formally each element of 
the Berry curvature tensor is itself an $n\times n$ matrix known as the non-Abelian
Berry curvature in the case of $n$ degenerate ground states
\cite{Shapere1989_1,Zanardi1999_1,Pachos1999_1}.  However, we use the fact
that the Berry curvature should not depend on the choice of ground state sector to
explicitly solve for only the case of positive magnetization along the $z$ direction.
We note that similar choices are employed in other analytic studies of the ground
state geometry of TFI chain \cite{Zhu2006_1,Kolodrubetz2013_2}.}. Then we wish to use QMC to measure the overlap
in \eref{eq:Fmunu_basic} as a function of $s \in (0,1)$:
\begin{widetext}
\be
\bra i \partial_\phi H \ket_\mathrm{asym} \equiv
\frac{\bra \psi(-v) | ( i \partial_\phi H ) | \psi(v) \ket}
{\bra \psi(-v) | \psi(v) \ket} \approx 
\frac{\bra \Rightarrow | e^{-H_M \delta \tau} e^{-H_{M-1} \delta \tau}\cdots 
e^{-H_{m} \delta \tau} \big[ i \partial_\phi H \big] e^{-H_{m-1} \delta \tau}
\cdots e^{-H_{1} \delta \tau} | \Uparrow \ket}
{\bra \Rightarrow | e^{-H_M \delta \tau} e^{-H_{M-1} \delta \tau}\cdots 
e^{-H_{m} \delta \tau} \big[ \mathbb 1 \big] e^{-H_{m-1} \delta \tau}
\cdots e^{-H_{1} \delta \tau} | \Uparrow \ket} \approx v F_{s\phi}~,
\label{eq:qmcmeas_simple}
\ee
where we have discretized the imaginary time evolution using an $M$-step Trotter decomposition
with time step $\delta \tau$. The velocity is
$v=\delta s / \delta \tau = 1/(M\delta \tau)$, while the Hamiltonians at each 
step are $H_p \equiv H\left(s=\frac{p-1/2}{M}\right)$.
To implement this ramp more easily, we use a trick from 
quasi-adiabatic QMC (QAQMC) and approximate \eref{eq:qmcmeas_simple} by
\cite{Liu2013_1}
\be
\bra i \partial_\phi H \ket_\mathrm{asym} \approx 
\frac{\bra \Rightarrow | (-H_M) (-H_{M-1})\cdots (-H_{m}) \big[ i \partial_\phi H \big] (-H_{m-1})
\cdots (-H_1) | \Uparrow \ket}{\bra \Rightarrow | (-H_M) (-H_{M-1})\cdots (-H_{m}) 
\big[ \mathbb{1} \big] (-H_{m-1}) \cdots (-H_1) | \Uparrow \ket}~,
\label{eq:qmcmeas}
\ee
\end{widetext}
where the effective ramp rate is now $v\approx -E_0/M$, with $E_0$ the ground state 
energy of $H$ at the measurement point $s_\mathrm{meas}=m/M$.  This expression becomes
exact in the limit of $M,E_0 \to \infty$, which can be achieved by taking the limit
of large system size $L$. 

In order to use an SSE-like method to extract
$F_{s\phi}$, we split the Hamiltonian up into bond and site operators (after adding a
constant offset):
\be
H(s,\phi=0) = -s \sum_j \big[ \mathbb{1}_j - \sigma^x_j \big] - (1-s)\sum_{<j j'>} 
\big[ \sigma^z_j \sigma^z_{j'} + \mathbb{1}_{j j'} \big] ~,
\label{eq:ops_ising}
\ee
as in equilibrium SSE-QMC simulations of the Ising model \cite{Sandvik2003_1}.
We also must introduce operators corresponding to the generalized force, 
\beq
\nn
\partial_\phi H \big| _{\phi=0} &=& i (1-s) \sum_{<j j'>} \left( \sigma^+_j \sigma^+_{j'} - 
\sigma^-_j \sigma^-_{j'} \right) 
\\&=& -\frac{1-s}{2} \sum_{<j j'>} \left[ \sigma^y_j \sigma^z_{j'} + \sigma^z_j \sigma^y_{j'} \right] ~.
\eeq
To this end, we define the measurement operator
\beq
\nn
\mathcal M &=& \sum_{<j j'>} \big[ \mathbb{1}_{jj'} - i \sigma^y_{j} \sigma^z_{j'}
- i \sigma^z_{j} \sigma^y_{j'}  + \sigma^x_{j} \sigma^x_{j'}  \big] 
\\ &=& N_\mathrm{bond} \mathbb{1} + \frac{2}{1-s} \big( i \partial_\phi H \big) + H_{xx} ~.
\label{eq:ops_meas}
\eeq
Here the identity term $\mathbb 1$ is used in sampling the denominator of 
\eref{eq:qmcmeas}, the generalized force term $i\partial_\phi H$ is used for the numerator,
and the spurious $H_{xx}$ term is
included to maximize ergodicity.
Finally, we measure the ground state energy $E_0$ within the same simulation by sampling
\be
\frac{\bra \Rightarrow | (-H_M) \cdots \big[ H_m \big] \cdots (-H_1)| \Uparrow \ket}
{\bra \Rightarrow | \cdots \big[ \mathbb{1} \big] \cdots | \Uparrow \ket} = E_0 + O(v^2)~.
\label{eq:Egs}
\ee

We efficiently sample the overlaps in Eqs.~\ref{eq:qmcmeas} and \ref{eq:Egs}
via cluster updates similar to those in conventional SSE-QMC \cite{Sandvik2003_1}; details of
the algorithm, including the cluster updates, 
can be found in the appendix.  In addition to standard diagonal and cluster
updates for the Ising Hamiltonians,
we introduce additional updates to sample the operators from $\mathcal M$.
The crucial idea there is that the $\sigma^y \sigma^z$ terms in $i\partial_\phi H$ are
``half-diagonal,'' i.e. $\sigma^y$ flips the spin and $\sigma^z$ does not.  
Therefore, for the cluster update, we treat the operators
in $\mathcal M$ as two separate sites, which are updated according to the same rules as
the on-site $\sigma^x$ and $\mathbb 1$ operators.  So, for example, if the incoming
vertex is to the $\sigma^z_{j'}$ vertex of a $\sigma^y_j \sigma^z_{j'}$ operator,
then the vertex is flipped to give $\sigma^x_j \sigma^x_{j'}$
\footnote{For the purposes of QMC sampling, $\sigma^x$ and $\sigma^y$ are
treated the same, since they both flip the spin with the same absolute values of
their matrix elements.  The difference enters in the sign of the measurement; see
appendix for details.}.

\begin{figure}
\includegraphics[width=.7 \linewidth]{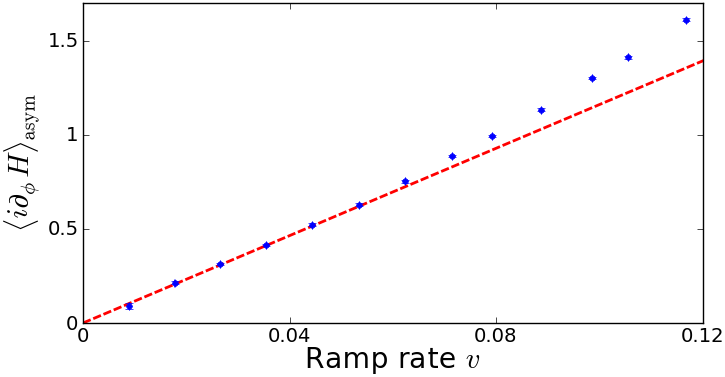}
\caption{To extract the Berry curvature of the one dimensional
TFI model from QMC, we plot the generalized force $\bra i 
\partial_\phi H \ket_\mathrm{asym}$ as a function of ramp rate $v_\mathrm{asym}$
for $s_\mathrm{meas}=0.25$ and $L=50$.
The slope at low velocity gives the Berry curvature $F_{s\phi}$, 
which matches well with the exact result (dashed line).}
\label{fig:tfi_berry_init}
\end{figure}

The simple method as described above works well for many cases, but devotes
unnecessary time to ramping through values of $s$ far from the measurement point.
We therefore improve the algorithm by starting
with $s_i$ and $s_f$ closer $s_\mathrm{meas}$,
zooming in on a range of small width $\Delta s = s_f - s_i$ around it
\footnote{For this paper, we specifically choose the symmetric case $s_i = s_\mathrm{meas}-\Delta s/2$
and $s_f=s_\mathrm{meas}+\Delta s / 2$}.
By changing $s_i$ and $s_f$, $\bra \Rightarrow |$ 
and $|\Uparrow \ket$ are no longer the ground states at the boundaries.
In principle, one could modify the algorithm to first project to
the ground state at each end; however, as we are performing imaginary time ramps, 
the dynamics will continually project toward the ground state.  Therefore, the boundary states don't matter
for slow enough ramps \cite{DeGrandi2011_1}, and our algorithm works as before
if $M$ is large enough to allow the initial
transients to relax \footnote{Indeed, with this modification the identity of the boundary states
no longer matters at all, and one could choose them as desired for numerical
convenience. For the remainder of this paper, we stick with the original boundary conditions
$|\Rightarrow\ket$ and $|\Uparrow\ket$ for convenience.}.

For a given $s_\mathrm{meas}$,
as the operators $(-H_p)$ are applied $M$ times, 
physical observables relax to their ground state value roughly exponentially \cite{Sandvik2010_1}
with decay rate proportional to the (many-body) energy gap.
Therefore, to approach the limit $M\to\infty$ where the algorithm becomes exact,
we measure $F_{s\phi}$ while varying $M$, and perform an 
exponential fit to estimate the decay constant $M_\mathrm{decay}$, and subsequently
work with $M$ much larger than $M_\mathrm{decay}$\footnote{We use roughly 
$M \approx 10 M_\mathrm{decay}$.}.

With this improvement, we now see that our QMC method scales similarly to 
conventional ground state methods. As noted earlier,
the number of steps required to reach the ground state scales as $M \sim L^d / \Delta$ \cite{Sandvik2010_1}.
Then, to obtain a given velocity, we simply tune the range $\Delta s$.  Therefore, the scaling of the
number of steps $M$ to get a given velocity is $M \sim \Delta^{-1}$, as opposed to $M \sim \Delta^{-2}$
as might be expected from more naive methods (to get $v \ll \Delta^2$). 
In addition, there is only one step (the measurement) which involves
signed sums, and this ``sign problem'' does not scale with 
system size.  Therefore, there is no exponentially bad sign problem;
indeed, since the scaling of $M$ is dominated by the number of steps required to
reach the ground state, it is identical to similar ground state algorithms.

\begin{figure}
\includegraphics[width=.7 \linewidth]{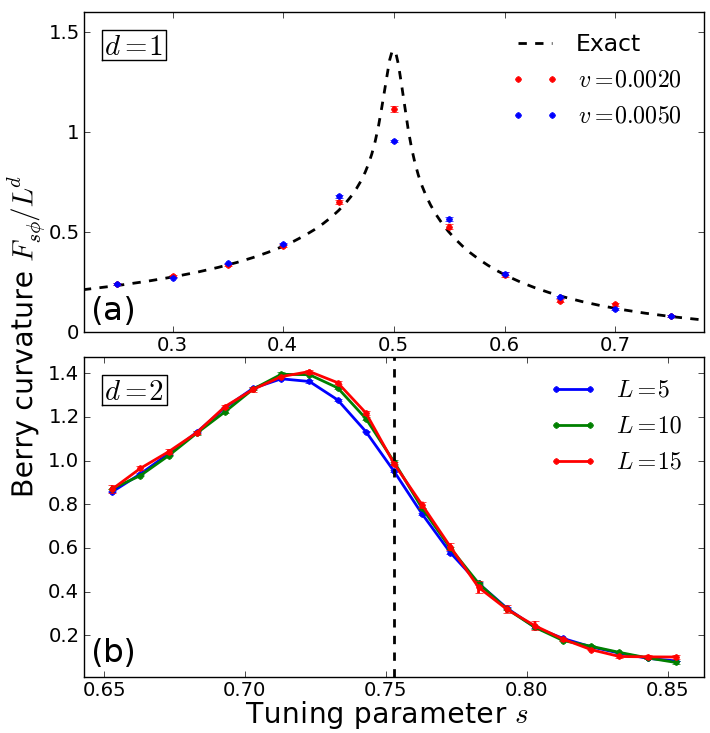}
\caption{Berry curvature of the TFI model in one (a) and two (b) dimensions.
(a) Results for one-dimensional model on an $L=50$ site lattice at
two small ramp rates.  The dashed line shows the exact result ($v \to 0$) for comparison.
(b) Measured Berry curvature of the $d=2$ TFI model as a function of system size at
fixed $v \sim L^d \Delta s / M = 10^{-2}$. Unlike the one-dimensional case, 
the Berry curvature does not diverge in the thermodynamic limit.}
\label{fig:tfi_berry_phasediag}
\end{figure}

\emph{Results -- }
For the one-dimensional
case, the TFI model is exactly solvable via Jordan-Wigner transformation \cite{Lieb1961_1}, from
which the Berry curvature can easily be obtained numerically \cite{Kolodrubetz2013_2}. 
As an initial test of the algorithm, we check that the exact and QMC solutions
match for an example point in \fref{fig:tfi_berry_init}.  As 
expected, the generalized force $i\partial_\phi H$ scales linearly with
$v$ at low ramp rates, with slope given by the Berry curvature.

We now proceed to obtain the Berry curvature of the full phase diagram in the 
one-dimensional model with $L=50$ and in two dimensions for a range of $L$
(\fref{fig:tfi_berry_phasediag}). The 1D results are compared to the exact
solution, where clear deviations are seen at 
finite velocity.  This is not surprising, given that the adiabatic limit
requires that the velocity be less than the gap squared: $v \ll \Delta^2$ 
\cite{DeGrandi2010_2}.  So, given that
the gap is vanishes at the critical point ($s=1/2$) in the thermodynamic limit,
the finite-velocity deviations are strongest near there \footnote{When
working with finite values of the velocity and system size, we can extend this
analysis by performing Kibble-Zurek scaling of the Berry curvature to 
near the critical point \cite{Liu2013_1}.}. However, for the case of 
a gapped spectrum, low velocities are already
sufficient to get the Berry curvature.

Having established that the method works in the integrable one-dimensional case, we can now easily
do the calculation for the non-integrable case of two dimensions.  
The results of this simulation are shown in \fref{fig:tfi_berry_phasediag}b.
We immediately see a qualitative difference from the one-dimensional model: 
the Berry curvature no longer is divergent (in the thermodynamic limit) at the quantum critical point.
Indeed, unlike in one dimension, the two-dimensional Berry curvature appears to approach
the thermodynamic limit for systems as small as $L=10$.

We can understand this difference using critical scaling arguments.
As shown in \rcite{Venuti2007_1}, the scaling dimension of the Berry curvature
is given by $F_{s\phi}/L^d \sim |s-s_c|^{-\nu_s (d+2z-\Delta_s-\Delta_\phi)}
\equiv |s-s_c|^\alpha$,
where $\Delta_\lambda = d + z - 1/\nu_\lambda$ is the scaling of the operator $\partial_\lambda H$. 
For the one-dimensional TFI model, we get $\alpha=0$ (i.e., $F_{s\phi}/L \sim \log(|s-s_c|)$),
while for two dimensions (using the exponents in \rcite{Liu2013_1}), we 
get $\alpha \approx 0.258$, which is singular but not divergent.  
Thus, the 2D Berry curvature will be some smooth function of $s$ plus a non-divergent singularity at the
critical point.  While our simulations are unable to resolve this singularity,
\fref{fig:tfi_berry_phasediag}b is consistent with its existence and clearly rules
out the existence of a divergent Berry curvature at the critical point.

\emph{Conclusions -- }
In conclusion, we have implemented a QMC method to measure Berry curvature.  Despite the fact that
Berry curvature measures properties of the ground state Berry phase, our algorithm has no
sign problem for the same set of Hamiltonians as more conventional QMC methods.  In addition,
the algorithm scales efficiently with both system size and energy gap. Using our specific 
implementation similar to quasi-adiabatic QMC, we solved for the ground state Berry curvature
of the transverse-field Ising model in one and two dimensions.  In agreement with critical
scaling theory, we saw a qualitative distinction between the models, namely
the presence (absence) of divergence in the Berry curvature in one (two) dimensions.

To our knowledge, this is the first demonstration of a QMC method to solve
for the Berry curvature with respect to global 
(as opposed to local \cite{Motoyama2013_1}) coupling parameters.
Having seen that this idea works for a simple case, the
idea is readily extensible to other models. Possible extensions include understanding the response
to twisted boundary conditions (``flux insertion'' \cite{Assaad1993_1}) and 
topological phase transitions in bosons \cite{Nightingale1999_1}, spins \cite{Syljuasen2002_1}, 
or even fermions \cite{Sorella1989_1}.  An interesting open question is whether a small
ramp along a direction with a sign problem is sufficient to create a sign problem
in the algorithm. If it is possible to do small ramps around a
sign-problem-free point, this would further open the class of problems solvable
via such a method.

\emph{Acknowledgments -- }
We thank Anatoli Polkovnikov and Anders Sandvik for crucial discussions and
Cheng-Wei Liu for sharing his QAQMC code.  We gratefully acknowledge financial support from
AFOSR grant number FA9550-10-1-0110 and NSF grant number PHY-1211284.

\bibliography{../../References/References}

\appendix
\subsection{Appendix: Details of QMC scheme}
\label{sec:qmc_details}

In this section, we detail the QMC scheme used to solve for the Berry
curvature of the TFI model.  With the exception of the measurement operator 
$\mathcal M$, it is very similar to the algorithm found in \rcite{Liu2013_1},
which in turn is based on SSE-QMC for the TFI model \cite{Sandvik2003_1}.  We
start by briefly reviewing the QMC update steps for the Hamiltonians, then
discuss how this is modified by the presence of $\mathcal M$. Finally, we discuss
the computation of observables, namely the generalized force $i\partial_\phi H$
and the ground state energy $E_0$.

In the SSE method, the Monte Carlo configuration consists of an operator string
with one operator for each step $p$. 
Throughout this appendix, we discuss the improved version of the algorithm,
in which $s_i = s_\mathrm{meas} - \frac{\Delta s}{2}$ and 
$s_f = s_\mathrm{meas} + \frac{\Delta s}{2}$.  Therefore, step $p$
samples from the Hamiltonian $H_p \equiv H\left(s=s_p\right)$, where
$s_p=s_i+\left( \frac{p-1/2}{M} \right) \Delta s$.  We denote by $H_{i_p}^p$
such an operator sampled from $H_p$, 
where $i_p$ iterates over the possible operators in \eref{eq:ops_ising}.
SSE generally requires sampling over spin states at the boundaries,
but as we'll see, this is unnecessary in the current case. We sample
an additional operator for the measurement, which we denote $\mathcal M_{i_\mathcal{M}}$.
So for our algorithm, a configuration $c$ is represented by the operator
string $(H^1_{i_1}, \ldots, H^M_{i_M}, \mathcal M_{i_{\mathcal M}})$.

Given the desired measurement (\eref{eq:qmcmeas}),
we define the sampling weight $w$ for configuration $c$ to be proportional to
\be
w(c) \propto \bra \Rightarrow | (-H^1_{i_1}) \cdots (-H^m_{i_m}) 
(\mathcal M_{i_{\mathcal M}}) \cdots (-H^M_{i_M}) | \Uparrow \ket ~.
\ee 
These weights can be negative because the measurement operator
has terms of the form $i\sigma^y \sigma^z$ which have negative
matrix elements.  Therefore, the sampling probability
is just the amplitude of this weight: $p(c) = |w(c)|$.  

Configurations can be efficiently sampled according to this probability
distribution using a straightforward extension of SSE cluster updates.
The basic QMC step of SSE-QMC on the TFI model consists of two parts: the
diagonal update, followed by the off-diagonal cluster update \cite{Sandvik2003_1}.
The diagonal update only re-samples operators that are diagonal in the spin basis.
Since we choose to quantize the spins along the $z$ axis, the diagonal operators
in \eref{eq:ops_ising} are the Ising operator and the heat bath operator $\mathbb 1_j$.  At each
step $p$ for which one of these operators is in the operator string (i.e., the 
current Monte Carlo configuration), a new diagonal operator is selected at random
such that the new configuration $c'$ is selected with probability proportional
to $p(c')$.  The heat bath positions are therefore selected at
random, while the Ising bond is only allowed to be inserted in positions where the
spins on either side of the bond are aligned.
The diagonal update of $\mathcal M$ is also straightforward, since the only diagonal
operator is $\mathbb{1}_{j j'}$.  Therefore, if the 
operator $\mathcal M_{i_\mathcal{M}}$ starts as an identity, we move it to a random bond
during the diagonal update step. 

The cluster update is more complicated, and is responsible for the efficiency of
the SSE algorithm.  The idea is to generate a connected cluster of ``nodes,'' where each node
is an entrance or an exit vertex from a given operator.  Then, with probability
$p=1/2$, each cluster is flipped, meaning that the spin state at each node is 
flipped with the operator changed accordingly.  The rules for generating the cluster
are simple: nodes that enter a heat bath ($\mathbb 1_j$) or 
spin flip ($\sigma^z_j$) operator terminate
the cluster, while nodes entering an Ising bond continue the cluster growth from the
other three nodes on that Ising operator \cite{Sandvik2003_1}.  
The $50\%$ acceptance rate comes from the fact that a cluster
has the same sampling probability $p(c)$ before and after flipping.  There is one slight
difference between our situation and that of some other algorithms: one of our boundaries
has all spins pointing up ($|\Uparrow \ket$), meaning that clusters that are in contact
with that boundary are not flipped.  This is in contrast to the other
boundary $|\Rightarrow \ket$, where clusters 
touching the boundary can be flipped because the $|\Rightarrow\ket$ state has equal
overlap with all spin states in $\uparrow/\downarrow$ basis.

In extending these ideas to the measurement step, given
by the operator $\mathcal M$ in \eref{eq:ops_meas}, the tricky part is that the 
$\sigma^y \sigma^z$ terms are ``half-diagonal.'' They are therefore similar in spirit to
two copies of the site operators $\sigma^x_j$ and $\mathbb{1}_j$. By introducing the 
otherwise useless $\sigma^x \sigma^x$ operator, which flips both sites,
we can complete this analogy by doing the cluster update of all terms in
$\mathcal M$ by just treating them as two separate site operators, which just
happen to lie on the same bond.  For example, if we start with operator
$\sigma^y_j \sigma^z_{j'}$ on some bond $<j j'>$ and generate a cluster that
comes in with entrance vertex on site $j'$, then we stop the cluster upon entering
the $\sigma^y \sigma^z$ vertex. If the cluster is subsequently chosen to be flipped, we
flip the operator at site $j'$ to get $\sigma^x_j \sigma^x_{j'}$.  As before, the
clusters can be flipped with probability $1/2$, because the amplitude $p(c)$ is
the same before and after flipping.  Note that, in flipping the operators,
$\sigma^y$ and $\sigma^x$ are treated as identical because they have the same
magnitude of their matrix elements (i.e., they give the same $p(c)$).  The
signs that distinguish these Pauli matrices enter into the actual measurement.

In taking operator expectation values (with respect to $p(c)$),
the sign of the weight $w(c)$ must be considered.  More explicitly, the
asymmetric expectation value of $i\partial_\phi H$ can be measured using the indicator observables
\beq
\mathcal O_{yz}(c) &=& \begin{cases}
\sgn \big[ w(c) \big] & \mbox{if } \mathcal M_{i_{\mathcal M}} \in \{
\sigma^z_j \sigma^y_{j'} \} \\
0 & \mbox{otherwise} 
\end{cases} \\
\mathcal O_{\mathbb 1} (c) & = & \begin{cases}
1 & \mbox{if } \mathcal M_{i_{\mathcal M}} \in \{
\mathbb 1_{j j'} \} \\
0 & \mbox{otherwise} 
\end{cases}
\eeq
Then the overlap is given by
\be
\bra i \partial_\phi H \ket_\mathrm{asym} = \frac{N_\mathrm{bond} (1-s)}{2} 
\frac{\bra \mathcal O_{yz} \ket_{p(c)}}
{\bra \mathcal O_{\mathbb 1} \ket_{p(c)}} ~,
\label{eq:Fsphi_qmc}
\ee
where $\bra \cdots \ket_{p(c)}$ is the statistical expectation value and
the $\frac{N_\mathrm{bond} (1-s)}{2}$ term comes from the prefactor
in \eref{eq:ops_meas}.

To obtain the velocity $v \approx -E_0 / M$, we need access to the ground
state energy $E_0$. Rather than separately solving this energy, we use the 
approximate form
\be
E(v) \equiv \frac{\bra \psi(-v) | H | \psi(v) \ket}{\bra \psi(-v) | \psi(v)},
\ee
which can be obtained at the same time as we measure the asymmetric expectation value
$\bra i \partial_\phi H \ket_\mathrm{asym}$. 
Furthermore, since this observable is diagonal in the energy basis,
the leading order correction will be of order $v^2$ \cite{DeGrandi2011_1}, so that
$E(v) = E_0 + O(v^2)$.  Therefore, in the limit $v \to 0$, using $E(v)$ 
in place of $E_0$ in the formula for $F_{s\phi}$ should still be accurate to 
order $v$.  

The energy consists of two terms: an Ising energy that is diagonal
in the $z$ basis and a field energy that is off-diagonal.  Diagonal and off-diagonal
operators are generally measured differently in SSE-QMC \cite{Syljuasen2002_1}.
In particular, while both energies are measured only for the case where
$\mathcal M_{i_M}$ is an identity matrix, the Ising energy is measured precisely at 
the measurement step $m$, while the field energy is averaged over
steps $m-1$ and $m$.  More explicitly, the indicator observable for the Ising energy is
\be
\mathcal O_\mathrm{ising}(c) = \begin{cases}
\sum_{<j j'>} \eta_{j j'} & \mbox{if } \mathcal M_{i_{\mathcal M}} 
 \in \{ \mathbb 1_{j j'} \} \\
0 & \mbox{otherwise} 
\end{cases} ~,
\ee
where $\eta_{j j'}=1$ if the spin states on sites $j$ and $j'$ match
in the configuration $c$ and 
$\eta_{j j'}=-1$ otherwise.  Similarly, the field energy at step
$p$ is given by the indicator variables
\beq
\mathcal O_\mathrm{field}^p (c) &=& \begin{cases}
1 & \mbox{if } \mathcal M_{i_{\mathcal M}} 
 \in \{ \mathbb 1_{j j'} \} \mbox{ and }  
H^p_{i_p} \in \{ \sigma^x_j \} \\
0 & \mbox{otherwise} 
\end{cases} \\
\mathcal O_\mathrm{bath}^p (c) &=& \begin{cases}
1 & \mbox{if } \mathcal M_{i_{\mathcal M}} 
 \in \{ \mathbb 1_{j j'} \} \mbox{ and }  
H^p_{i_p} \in \{ \mathbb{1}_j \} \\
0 & \mbox{otherwise} 
\end{cases} ~.
\eeq
Then the Ising and field energy densities at the point $s_\mathrm{meas}$ are
\be
\frac{E_\mathrm{ising}}{N_\mathrm{site}} = \frac{J_z \bra \mathcal O_\mathrm{ising} \ket_{p(c)}}
{N_\mathrm{bond} \bra \mathcal O_{\mathbb 1} \ket_{p(c)}} ~~~,~~~
\frac{E_\mathrm{field}}{N_\mathrm{site}} = \frac{h \bra \frac{\mathcal O_\mathrm{field}^{m-1} 
+ \mathcal O_\mathrm{field}^{m}}{2} \ket_{p(c)}}
{\bra \mathcal O_\mathrm{bath} \ket_{p(c)}} ~,
\ee
in terms of which the ground state energy is well approximated by (see \eref{eq:ops_ising})
\be
E_M \equiv E_\mathrm{ising} + E_\mathrm{bond} + N_\mathrm{site} ~.
\ee

In summary, we use the following formula as a QMC estimate 
of the Berry curvature,
\be
F_{s \phi}(M) \equiv \frac{-M N_\mathrm{bond} (1-s) \bra \mathcal O_{yz} 
\ket_{p(c)}}{2 E_M \bra \mathcal O_{\mathbb 1} \ket_{p(c)}} ~,
\label{eq:Fsphi_finite_vel}
\ee
which becomes exact in the limit $M \to \infty$.

\end{document}